\documentclass[aps, superscriptaddress,12pt]{revtex4-1}
\usepackage[colorlinks=true,allcolors=blue!92!black]{hyperref}
\usepackage{amssymb}
\usepackage{amsmath}
\usepackage{subfigure}
\usepackage{amsfonts}
\usepackage{xcolor}
\usepackage{epsfig}
\usepackage{color}
\usepackage{bbm}
\usepackage{bm}
\usepackage{tabularx}
\usepackage{multirow}
\usepackage{mathtools}
\usepackage{xfrac}
\usepackage{amsfonts}
\usepackage{latexsym}
\usepackage{amsmath,amssymb}
\usepackage{mathrsfs}
\usepackage{setspace}
\usepackage{color,xcolor}
\usepackage{bm}
\usepackage{slashed}
\usepackage{url}
\usepackage{empheq}
\usepackage{textcomp}
\usepackage{dcolumn}
\usepackage{ulem}
\usepackage{cancel}
\usepackage{braket}
\usepackage[title]{appendix}
\usepackage{lipsum}
\usepackage{ulem}
\usepackage{cancel}
\usepackage{color}
\usepackage{cprotect}

\numberwithin{equation}{section}

\def \l{\left}
\def \r{\right}
\def\beq{\begin{equation}}
	\def\eeq{\end{equation}}
\def\bea{\begin{eqnarray}}
	\def\eea{\end{eqnarray}}

\newcommand{\q}{\mathbf{q }}
\newcommand{\p}{\mathbf{ p}}
\newcommand{\x}{\mathbf{ x}}
\def\k{\textbf{k}}



\begin{document}



\title{ \Large Induced Circular Polarization on Photons Due to Interaction with Axion-Like Particles in Rotating Magnetic Field of Neutron Stars}

	\author{M. Sharifian}
	\email[]{mohammadsharifian.physics@gmail.com}
\affiliation{Department of Physics, Isfahan University of Technology, Isfahan 84156-83111, Iran}
\affiliation{ICRANet-Isfahan, Isfahan University of Technology, 84156-83111, Iran}

\author{M. Zarei}
	\email[]{m.zarei@iut.ac.ir}
\affiliation{Department of Physics, Isfahan University of Technology, Isfahan 84156-83111, Iran}
\affiliation{ICRANet-Isfahan, Isfahan University of Technology, 84156-83111, Iran}

\author{S. Matarrese}
	\email[]{sabino.matarrese@pd.infn.it}
\affiliation{Dipartimento di Fisica e Astronomia “Galileo Galilei” Universita` di Padova, 35131 Padova, Italy}
\affiliation{INFN, Sezione di Padova, 35131 Padova, Italy}
\affiliation{INAF - Osservatorio Astronomico di Padova, I-35122 Padova, Italy}
\affiliation{Gran Sasso Science Institute, I-67100 L'Aquila, Italy}

\author{R. Turolla}
\email[]{turolla@pd.infn.it}
\affiliation{Dipartimento di Fisica e Astronomia “Galileo Galilei” Universita` di Padova, 35131 Padova, Italy}
\affiliation{Mullard Space Science Laboratory, University College London, Holmbury St. Mary, Dorking, Surrey RH5 6NT, UK}


\begin{abstract}
We investigate how the photon polarization is affected by the interaction with axion-like particles (ALPs) in the rotating magnetic field of a neutron star (NS). Using quantum Boltzmann equations the study demonstrates that the periodic magnetic field of millisecond NSs enhances the interaction of photons with ALPs and creates a circular polarization on them.  A binary system including an NS and a companion star could serve as a probe. When the NS is in front of the companion star with respect to the earth observer, there is a circular polarization on the previously linearly polarized photons as a result of the interaction with ALPs there. After a half-binary period, the companion star passes in front of the NS, and the circular polarization of photons disappears and changes to linear. The excluded parameter space for a millisecond NS with 300~Hz rotating frequency, highlights the coupling constant of $1.7\times10^{-11}~\text{GeV}^{-1}\leq g_{a\gamma\gamma}\leq1.6\times10^{-3}~\text{GeV}^{-1}$ for the  ALP masses in the range of $7\times10^{-12}~\text{eV}\leq m_a\leq1.5\times 10^{3}~\text{eV}$.  
\end{abstract}

\maketitle


\pagestyle{plain}
\setcounter{page}{1}
\newcounter{bean}

\setcounter{tocdepth}{2}

\newpage
\tableofcontents

\section{Introduction}

Axions and ALPs are massive particles that arise in theories related to CP-violation in the standard model \cite{peccei1977constraints,wilczek1978problem,Kim1979weak,Dine1981a}, string theory \cite{svrcek2006axions,marsh2016axion}, and supersymmetry \cite{dobrescu2000light}. Currently, ALPs are considered to be one of the leading candidates for dark matter \cite{chadha2022axion,Dine1983the}. Numerous astrophysical probes have been proposed to study ALPs, exploiting either their gravitational effects \cite{arvanitaki2011exploring,baryakhtar2021black} or their interactions with particles from the standard model \cite{Rodd2018Listening}. The ALP-photon interaction forms the basis of the majority of astrophysical ALP probes, benefitting from practical laboratory instruments capable of manipulating light. One approach involves investigating X-ray and gamma-ray emissions resulting from the conversion of ALPs in the presence of high magnetic fields surrounding white dwarfs and NSs \cite{dessert2019x,buschmann2021axion,Fortin2018constraining,Fortin2019x,Shakeri2017nonlinear}. Another astronomical probe is the search for ALP induced polarization in magnetic white dwarfs. The unpolarized photons leaving the white dwarf acquire polarization as a result of the ALP-photon interaction, indicating the presence of ALPs \cite{dessert2022upper}.

The center of a collapsing star is a good place to produce weakly interacting particles like ALPs \cite{Raffelt1996stars,Lucente2020}. There are many proposals for using supernova 1987A (SN 1987A) as a probe for ALPs in this context \cite{Giannotti2011New,Muller2023Invest,Diamond2023Axion}. The measurement of the neutrino burst from SN 1987A detected by Baksan, IMB, and Kamiokande II set constraints on the ALPs \cite{Burrows1987Neutrinos}. These restrictions are predicated on the idea that the rate of energy loss due to ALPs shouldn't be greater than the rate of energy loss due to neutrinos. SN 1987A bound on ALPs has an upper limit which is because the collapsing core is so dense that in the high couplings, the star is opaque to the radiating ALPs and traps them \cite{Lee2018Revisiting,Croon2021supernova}. The delayed and diffuse burst of gamma rays due to the decayed ALPs from SNe is another usage of SNe to probe ALP dark matter \cite{Jaeckel2018Decay}.

Using data from \textit{Fermi} Large Area Telescope (LAT) a closed region of parameter space can be excluded which is based on the hypothesis of ALP-photon interactions in an environment of galactic magnetic fields \cite{Meyer2016Fermi}. The focus of \textit{Fermi}-LAT is on the spectrum of the radio galaxy NGC 1275 to find irregularities in its data \cite{Ajello2016Search}. The common feature of some of astrophysical probes (e.g. \textit{Fermi}-LAT NGC 1275 and SN 1987A) is that there is an upper limit on the excluded area of them.

Significant progress in quantum technology has opened avenues for utilizing it in various instruments to investigate ALP interactions. These include spin-precession instruments \cite{dror2023sensitivity}, electromagnetic squeezing \cite{Malnou2019squeezed,zarei2022probing,sharifian2023probing}, Stern-Gerlach interferometers \cite{hajebrahimi2023axion}, and spin-based amplifiers \cite{su2021search,Sushkov2023quantum}. It has been shown that the ALP-photon interaction is enhanced in the presence of a magnetic field \cite{ouellet2019first,andriamonje2007improved,nitta2023arxiv,Shakeri2023probing}. This effect is further enhanced when employing either a space-periodic magnetic field \cite{zarei2022probing} or a time-periodic one \cite{sharifian2023probing,Seong2023Axion}.

In this study, we explore how the rotating magnetic field of an NS induces a transformation of previously linearly polarized photons into circularly polarized ones. Additionally, the temporal coherence of the magnetic field enhances this polarization effect, allowing for the detection of weak ALP-photon couplings within the exclusion region. We propose to observe this phenomenon in a binary system consisting of an NS and a companion star (CS). For Earth observers, the circular polarization becomes measurable when the CS undergoes partial eclipses by the NS. During such events, the photons emitted from the CS approach the surface of the NS closely, sensing its strong magnetic field. As the CS emerges from behind the NS, the distant observers on Earth can measure the circular polarization of the photons. However, this circular polarization is absent when the CS comes in front of the NS, as the photons emitted in such situations do not experience a sufficiently strong magnetic field from the NS.

The manuscript is organized as follows. In Section \ref{sec:stokesevol} we use Quantum Boltzmann Equation (QBE) to show how Stokes parameters of the photon evolve in the second order interaction with ALPs. We specifically study this time evolution near a NS in Section \ref{sec:NSStokes} and obtain the explicit differential equations of Stokes parameters. Section \ref{sec:scheme} is devoted to the results and proposing appropriate candidates to see the effect.

\section{time evolution of the Photon Stokes Parameters} \label{sec:stokesevol}
We start by describing the intensity and polarization of the photons emitted by the CS. The polarization matrix of the photons, which is expressed in terms of Stokes parameters, is given by
\begin{eqnarray} 
	\rho^{(\gamma)}_{ij}=\frac{1}{2}\left(\begin{array}{cc}I+Q & U-iV \\U+iV & I-Q\end{array}\right)~,
\end{eqnarray}
where $I$ represents the radiation intensity, $Q$ and $U$ parameterize the linear polarization, and $V$ denotes the circular polarization. Among these parameters, $I$ is always positive, while the other three parameters can have either positive or negative values. 

The time evolution of the Stokes parameters is determined by QBE. In the case of Markovian processes, where the influence of the environment on the system is negligible, the QBE can be formulated as a Master Equation. Consequently, the time evolution of the density matrix elements can be described by a set of differential equations, as follows \cite{Zarei:2021dpb}
\begin{equation}\label{boltzmann0}
	(2\pi)^3 k^0 \delta^3(0)\frac{d}{dt}\rho^{(\gamma)}_{ij}(t)=i\braket{\l[H_{\textrm{int}}(t),\hat{\mathcal{D}}_{ij}(\mathbf{k})\r]}-\int_{0}^{t} dt' \braket{\l[H_{\textrm{int}}(t),\l[H_{\textrm{int}}^\dag(t'),\hat{\mathcal{D}}_{ij}(\mathbf{k})\r]\r]}~,
\end{equation}
where $\hat{\mathcal{D}}_{ij}(\mathbf{k})=a^\dag_i(\mathbf{k}) a_j(\mathbf{k})$ is the photon number operator with the expectation value given by
\begin{equation}
	\l<\hat{\mathcal{D}}_{ij}(\mathbf{k})\r> = \text{Tr}(\hat{\rho}\hat{\mathcal{D}}_{ij}(\mathbf{k}))=(2\pi)^3 \delta^3(0) \rho^{(\gamma)}_{ij}(\mathbf{k})~,
\end{equation}
and $H_{\textrm{int}}$ is an effective interaction Hamiltonian describing the physical process and is determined using S-matrix element. The first term on the right-hand side of Eq. \eqref{boltzmann0} is referred to as the forward scattering term, while the second term is referred to as the damping or non-forward scattering term or conventional collision term. 
The following Lagrangian describes the axion-photon interaction
\bea
\mathcal{L}_{ a\gamma\gamma}=\frac{1}{4}\,g_{ a\gamma\gamma}\phi F_{\mu\nu}\widetilde{F}^{\mu\nu}~,
\eea
where $\phi$ is the ALP field, $F_{\mu\nu}$ is the electromagnetic field tensor with $\widetilde{F}^{\mu\nu}=\frac{1}{2} \epsilon^{\mu\nu\rho\sigma}F_{\rho\sigma}$ as its dual, and $g_{a\gamma\gamma}$ is the coupling constant. Since we are interested in an ALP-mediated process in an external magnetic field, we split the tensor $F_{\mu\nu}$ into a homogeneous background $\bar{F}_{\mu\nu}$ and a fluctuating quantum part $f_{\mu\nu}=\partial_{\mu}A_{\nu}-\partial_{\nu}A_{\mu}$ with $A_{\mu}$ denoting the photon field.
Therefore, by expanding the interaction term and ignoring its completely static and completely fluctuating components, the following. Lagrangian is resulted
\bea
\mathcal{L}_{a\gamma\gamma}=\frac{1}{2}g_{a\gamma\gamma}\phi \,\epsilon^{\mu\nu\rho\sigma} \bar{F}_{\mu\nu}\partial_{\rho}A_{\sigma}~.
\eea
This interaction Lagrangian allows us to investigate the interaction of a propagating polarized photon with a time-dependent background magnetic field through the intermediate ALP field.
\begin{figure}[t]	
	\includegraphics[width=.6\textwidth]{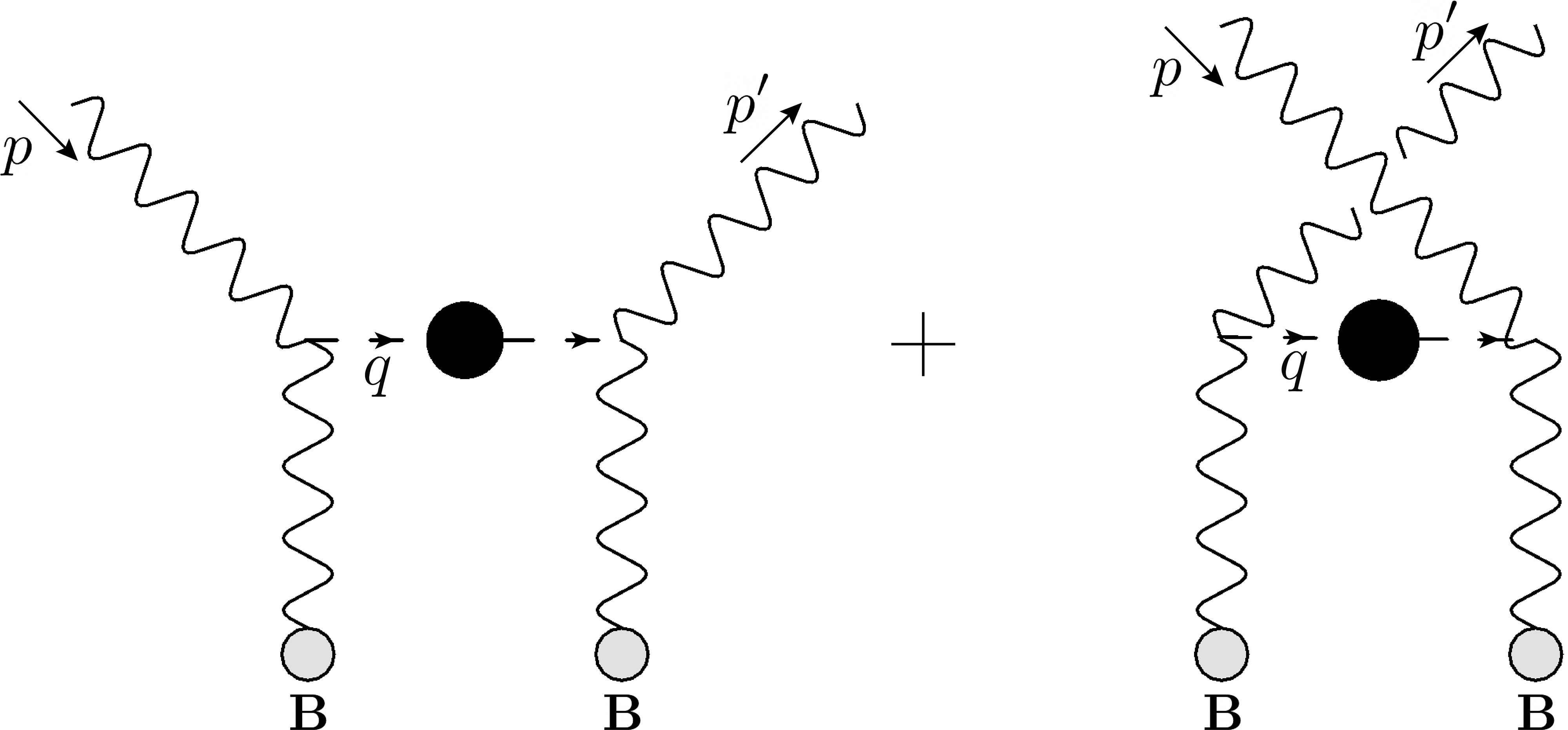}\centering
	\caption{Feynman diagrams depicting the scattering of photons with a background magnetic field mediated by off-shell ALPs.}
	\label{fig:feynmans}
\end{figure}
Fig.~\ref{fig:feynmans} illustrates the two Feynman diagrams associated with such process. The effective interaction Hamiltonian is found through the S-matrix operator associated with this process \cite{Zarei:2021dpb, kosowsky1995cosmic}
\bea
S^{(2)}(\gamma_s\rightarrow\gamma_{s'})=-i\int dt \,H_{\textrm{int}}(t)~,
\eea
where $H_{\textrm{int}}(t)$ describing the process shown in Fig.~\ref{fig:feynmans} is given by
\bea  \label{Hint1}
H_{\textrm{int}}(t)&=&\frac{g^2_{ a\gamma\gamma}}{8}\int d^4x' d^3x   \epsilon^{\mu\nu\rho\sigma}\epsilon^{\mu'\nu'\rho'\sigma'} \bar{F}_{\rho\sigma} (x)\bar{F}_{\rho'\sigma'} (x')D_F(x-x')
\left [\partial_{\mu}A^{-}_{\nu}(x) \partial_{\mu'}A^{+}_{\nu'}(x') 
\right.	\nonumber \\  && \left. ~~~~~~~~~~~~+~
\partial_{\mu'}A^{-}_{\nu'}(x') \partial_{\mu}A^{+}_{\nu}(x)\right ]~.
\eea
Here, $D_F(x-x')$ is the Breit-Wigner Feynman propagator for the ALP, and $A^{+}_{\mu}$ ($A^{-}_{\mu}$) is the electromagnetic field linear in the absorption (creation) operators of the photons 
\begin{eqnarray}      \label{Amu}
	A_\mu(x)=A_\mu^+(x)+A_\mu^-(x)~,
\end{eqnarray} 
where 
\begin{eqnarray}      \label{A+}
	A_\mu^+(x)= \int \frac{d^3 p}{(2\pi)^32p^0}\sum_s a_s(\p)\epsilon_s^\mu(\p)e^{-ip\cdot x}~,
\end{eqnarray}
and
\begin{eqnarray}      \label{A-}
	A_\mu^-(x)=\int \frac{d^3 p}{(2\pi)^32p^0}\sum_s a^\dagger_s(\p)\epsilon_s^{\mu*}(\p)e^{ip\cdot x}~,
\end{eqnarray}
where $\epsilon_s^{\mu}(\p)$ is the photon polarization and $a_s(\p)$ ($a^\dagger_s(\p)$) is the annihilation (creation) operator for photons obeying the canonical commutation relation 
\beq \label{modefunction}
\l[a_s(\p),a^{\dagger}_{s'}(\p')\r]=(2\pi)^32p^0\delta^3(\p-\p')~.
\eeq
Inserting the Fourier transforms \eqref{A+} and \eqref{A-} into Eq.~\eqref{Hint1}, we find
\begin{eqnarray} \label{H}
	H_{\textrm{int}}(t)&=&\frac{g^2_{a\gamma\gamma}}{2 }
	\,\sum_{s,s'}
	\int d^3rd^3r'dt' d\p d\p' \,\frac{d^4q}{(2\pi)^4}\omega_{\p}\omega_{\p'}D_F(q)
	\left[\bm{\varepsilon}_{s}(\p)\cdot\textbf{B}(\mathbf{r},t)\right]\left[\bm{\varepsilon}_{s'}^*(\p')\cdot\textbf{B}(\mathbf{r}',t')\right]
	\nonumber \\ &\times & 
	\left[
	e^{iq\cdot x' }
	e^{ip'\cdot   x}
	e^{-iq\cdot x }
	e^{-i p\cdot   x'}+ 
	e^{iq\cdot x' }
	e^{i p'\cdot   x'}
	e^{-iq\cdot x }
	e^{-i p\cdot   x}
	\right]a_{s'}^{\dagger}(\p')a_{s}(\p)~,
\end{eqnarray}
where $D_F(q)$ is the Fourier transform of $D_F(x)$, which is defined by 
\begin{eqnarray} \label{Dx1}
	D_F(x-x')=\int \frac{d^4q}{(2\pi)^4}\frac{1}{q^2-m_a^2+iq^0\Gamma_B(q^0)}e^{-iq\cdot (x-x')}~,
\end{eqnarray}
in which $m_a$ is the ALP physical mass, and $\Gamma_B(q^0)$ is the ALP decay rate in the presence of a time-dependent magnetic field and has been derived in Appendix \ref{appendix:decayrate}. Using the interaction Hamiltonian of Eq.~\eqref{H} in the QBEs of Eq.~\eqref{boltzmann0} we get the time evolution of the Stokes parameters. The details are presented in Appendix~\ref{appendix:stoke} which yields the following relation for Q parameter
\begin{eqnarray}\label{eq:qdot}
	&&(2\pi)^3 \delta^3(0)\dot{Q}(t)\nonumber\\
	&=&\frac{g^2_{a\gamma\gamma}}{2 }
	\,
	\int d^3rd^3r'dt'  \,\frac{d^4q}{(2\pi)^4}\,\omega_{\k}\,D_F(q)
	\nonumber \\ &\times & 
	\left[
	e^{-i(\q-\k)\cdot \mathbf{r}'}
	e^{-i (\k-\q)\cdot   \mathbf{r}}
	e^{-i(-q^0+\omega_{\mathbf{k}})t'}
	e^{-i(q^0-\omega_{\mathbf{k}'})t}+e^{-i(\q+\k)\cdot \mathbf{r}'}
	e^{-i(-\q-\k)\cdot \mathbf{r} }
	e^{-i(-q^0-\omega_{\k})t'}
	e^{-i(q^0+\omega_{\k})t}
	\right]\nonumber\\
	&\times&\Bigg[U(t)\Big(\left[\bm{\varepsilon}_{1}(\k)\cdot\textbf{B}(\mathbf{r},t)\right]
	\left[\bm{\varepsilon}_2^*(\k)\cdot\textbf{B}(\mathbf{r}',t')\right]-\left[\bm{\varepsilon}_2(\k)\cdot\textbf{B}(\mathbf{r},t)\right]\left[\bm{\varepsilon}^*_{1}(\k)\cdot\textbf{B}(\mathbf{r}',t')\right]
	\Big)~,\nonumber\\
	&+&iV(t)\Big(\left[\bm{\varepsilon}_{1}(\k)\cdot\textbf{B}(\mathbf{r},t)\right]
	\left[\bm{\varepsilon}_2^*(\k)\cdot\textbf{B}(\mathbf{r}',t')\right]+
	\left[\bm{\varepsilon}_2(\k)\cdot\textbf{B}(\mathbf{r},t)\right]\left[\bm{\varepsilon}^*_{1}(\k)\cdot\textbf{B}(\mathbf{r}',t')\right]\Big)\Bigg]~,\nonumber\\
\end{eqnarray}
and for U parameter
\begin{eqnarray}\label{eq:udot}
	&&(2\pi)^3 \delta^3(0)\dot{U}(t)\nonumber\\
	&=&\frac{g^2_{a\gamma\gamma}}{2 }
	\,
	\int d^3rd^3r'dt'  \,\frac{d^4q}{(2\pi)^4}\,\omega_{\k}D_F(q)
	\nonumber \\ &\times & 
	\left[
	e^{-i(\q-\k)\cdot \mathbf{r}'}
	e^{-i (\k-\q)\cdot   \mathbf{r}}
	e^{-i(-q^0+\omega_{\mathbf{k}})t'}
	e^{-i(q^0-\omega_{\mathbf{k}'})t}+e^{-i(\q+\k)\cdot \mathbf{r}'}
	e^{-i(-\q-\k)\cdot \mathbf{r} }
	e^{-i(-q^0-\omega_{\k})t'}
	e^{-i(q^0+\omega_{\k})t}
	\right]\nonumber\\
	&\times&\Bigg[iV(t)\Big(-\left[\bm{\varepsilon}_{1}(\k)\cdot\textbf{B}(\mathbf{r},t)\right]
	\left[\bm{\varepsilon}_1^*(\k)\cdot\textbf{B}(\mathbf{r}',t')\right]+\left[\bm{\varepsilon}_{2}(\k)\cdot\textbf{B}(\mathbf{r},t)\right]
	\left[\bm{\varepsilon}_2^*(\k)\cdot\textbf{B}(\mathbf{r}',t')\right]\Big)~,\nonumber\\
	&+&Q(t)\Big(-\left[\bm{\varepsilon}_{1}(\k)\cdot\textbf{B}(\mathbf{r},t)\right]
	\left[\bm{\varepsilon}_2^*(\k)\cdot\textbf{B}(\mathbf{r}',t')\right]+\left[\bm{\varepsilon}_{2}(\k)\cdot\textbf{B}(\mathbf{r},t)\right]
	\left[\bm{\varepsilon}_1^*(\k)\cdot\textbf{B}(\mathbf{r}',t')\right]\Big)\Bigg]~,\nonumber\\
\end{eqnarray}
and also V parameter
\begin{eqnarray}\label{eq:vdot}
	&&i(2\pi)^3 \delta^3(0)\dot{V}(t)\nonumber\\
	&=&\frac{g^2_{a\gamma\gamma}}{2 }
	\,
	\int d^3rd^3r'dt'  \,\frac{d^4q}{(2\pi)^4}\,\omega_{\k}\,D_F(q)
	\nonumber \\ &\times & 
	\left[
	e^{-i(\q-\k)\cdot \mathbf{r}'}
	e^{-i (\k-\q)\cdot   \mathbf{r}}
	e^{-i(-q^0+\omega_{\mathbf{k}})t'}
	e^{-i(q^0-\omega_{\mathbf{k}'})t}+e^{-i(\q+\k)\cdot \mathbf{r}'}
	e^{-i(-\q-\k)\cdot \mathbf{r} }
	e^{-i(-q^0-\omega_{\k})t'}
	e^{-i(q^0+\omega_{\k})t}
	\right]\nonumber\\
	&\times&\Bigg[U(t)\Big(-\left[\bm{\varepsilon}_{1}(\k)\cdot\textbf{B}(\mathbf{r},t)\right]
	\left[\bm{\varepsilon}_1^*(\k)\cdot\textbf{B}(\mathbf{r}',t')\right]+\left[\bm{\varepsilon}_{2}(\k)\cdot\textbf{B}(\mathbf{r},t)\right]
	\left[\bm{\varepsilon}_2^*(\k)\cdot\textbf{B}(\mathbf{r}',t')\right]\Big)~,\nonumber\\
	&+&Q(t)\Big(\left[\bm{\varepsilon}_{1}(\k)\cdot\textbf{B}(\mathbf{r},t)\right]
	\left[\bm{\varepsilon}_2^*(\k)\cdot\textbf{B}(\mathbf{r}',t')\right]+\left[\bm{\varepsilon}_{2}(\k)\cdot\textbf{B}(\mathbf{r},t)\right]
	\left[\bm{\varepsilon}_1^*(\k)\cdot\textbf{B}(\mathbf{r}',t')\right]\Big)\Bigg]~.\nonumber\\
\end{eqnarray}
Thus, there are three coupled differential equations for three Stokes parameters, demonstrating how each parameter can affect the others. In addition, there are energy-momentum and space-time integrals which should be done based on the magnetic field configuration. In these coupled differential equations, a time-periodic magnetic field can produce poles in the frequency integrals that create an amplification effect if they overlap with the propagator ones. The magnetic field profile of the NS is used in the following section to examine how the interaction with ALPs affects photons close to the NS.


\begin{figure}[t]
	\vspace{-1cm}
	\includegraphics[height=10cm]{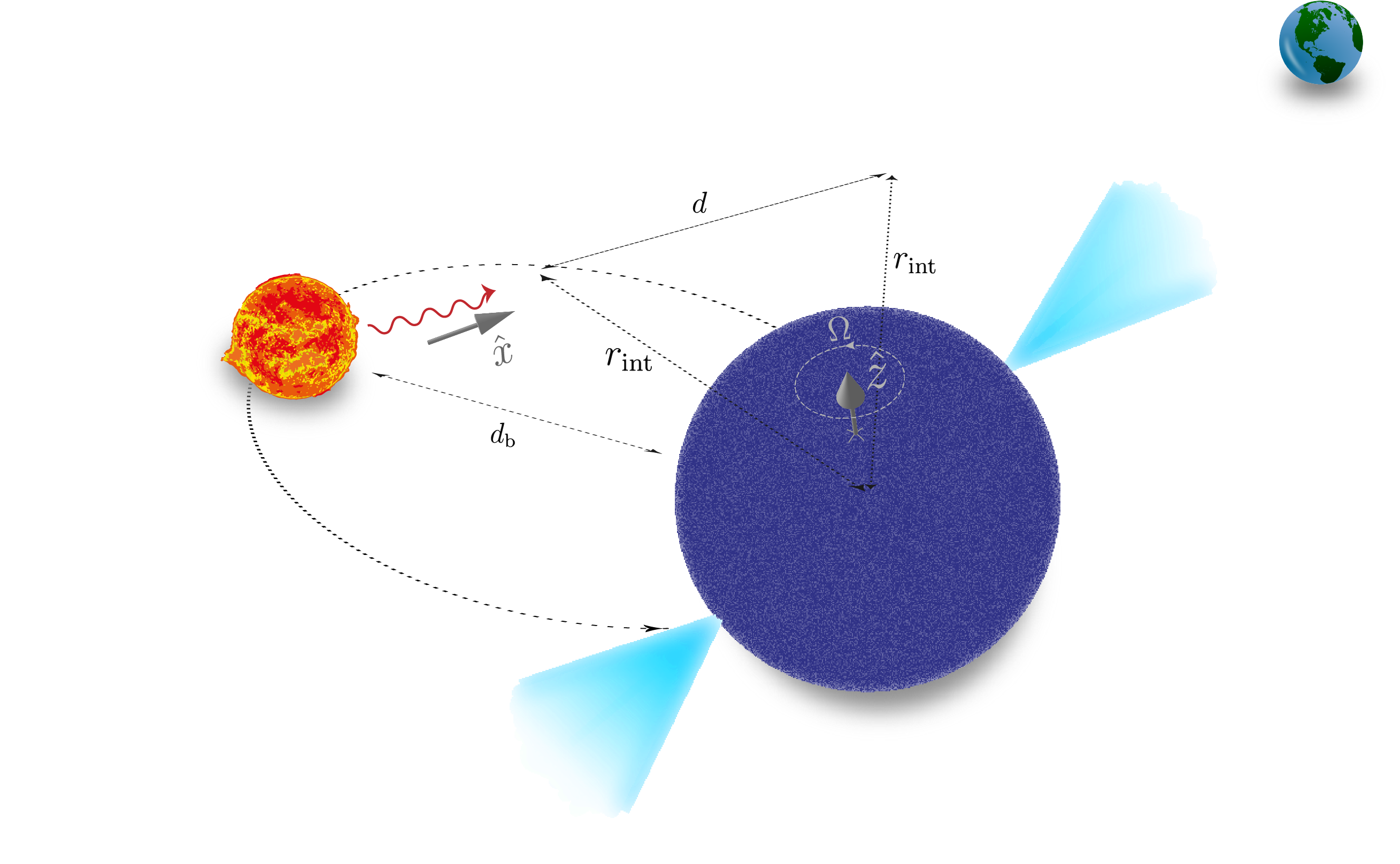}\centering
	\caption{A binary system consisting of an NS and a CS. The plot illustrates the effect of the NS's rotating magnetic field on the polarization of photons emitted by the CS as observed from Earth during partial eclipses. Circular polarization is detected when the CS approaches the NS closely, sensing its strong magnetic field, but is absent when the CS passes in front of the NS due to the weaker magnetic influence.}
	\label{fig:experiment}
\end{figure}

\section{The NS Effect on the stokes parameters}\label{sec:NSStokes}

Understanding the intriguing phenomena of induced circular polarization and its modulation in the presence of a rotating NS's magnetic field forms the basis of this section. We will delve into how the temporal coherence of the magnetic field plays a crucial role in enhancing this effect, allowing for the detection of remarkably weak ALP-photon couplings within the excluded area. Moreover, we propose to investigate this phenomenon further within the context of a binary system, comprising the NS and a CS. In Fig. \ref{fig:experiment}, we present a plot illustrating the phenomenon described above, showcasing the transformation of previously linearly polarized photons into circularly polarized ones under the influence of the rotating neutron star's magnetic field. The temporal coherence of the magnetic field, along with the binary system configuration, plays a crucial role in this observed effect.

\begin{figure}[t]
	\hspace{-2cm}
	\centering
	\includegraphics[width=.8\textwidth]{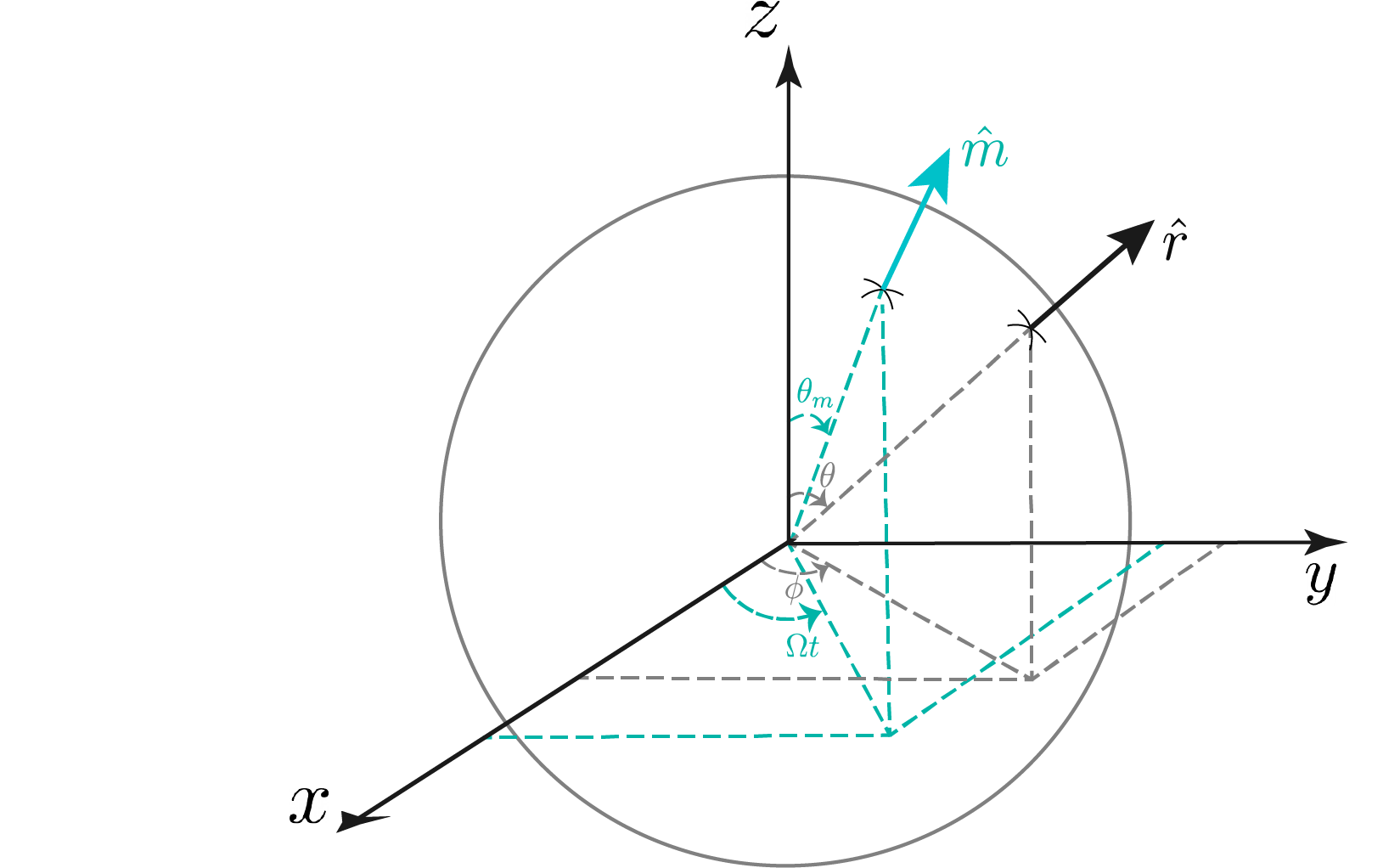}
	\vspace{.5cm}
	\caption{The spherical coordinate system used in the text with an origin on the center of the NS. The coordinate system's  radial unit vector, polar angle, and azimuthal angle are $\hat{\mathbf{r}}$, $\theta$, and $\phi$, respectively. The NS's dipole unit vector, misalignment angle of dipole, and rotating frequency are $\hat{\mathbf{m}}$, $\theta_m$, and $\Omega$, respectively. The NS rotation axis is in the $z$-direction. }\label{fig:coordinate}
	\vspace{-3cm}
\end{figure}

In the subsequent discussion, we will introduce and examine the magnetic field generated by neutron stars, outlining its specific configuration that underlies the phenomena we have been describing.
The magnetic field of the neutron star (NS) is typically assumed to adopt a dipole form~\cite{hook2018radio}, 
\begin{eqnarray}\label{bg}
	\textbf{B}(\mathbf{r},t)=\frac{B_0}{2}\left(\frac{r_0}{r}\right)^3\big[3(\hat{\mathbf{m}}\cdot\hat{\mathbf{r}})\hat{\mathbf{r}}-\hat{\mathbf{m}}\big]~,
\end{eqnarray}
where $r_0$ denotes the NS radius, $B_0$ represents its magnetic field strength near the surface, $\hat{\mathbf{r}}$ denotes the radial unit vector, and $\hat{\mathbf{m}}$ signifies the dipole unit vector. As shown in Fig.~\ref{fig:coordinate}, in the Cartesian coordinate system $\hat{\mathbf{m}}$ and $\hat{\mathbf{r}}$ are written as follows
\begin{eqnarray}
	\hat{\mathbf{m}}&=&\sin(\theta_m)\cos(\Omega t)\,\hat{\mathbf{x}}+\sin(\theta_m)\sin(\Omega t)\,\hat{\mathbf{y}}+\cos(\theta_m)\,\hat{\mathbf{z}},\nonumber\\
	\hat{\mathbf{r}}&=&\sin(\theta)\cos(\phi)\,\hat{\mathbf{x}}+\sin(\theta)\sin(\phi)\,\hat{\mathbf{y}}+\cos(\theta)\,\hat{\mathbf{z}}~,
\end{eqnarray}
where $\theta_m$ represents the misalignment angle of the NS dipole, $\Omega$ denotes the rotation frequency of the NS, $\theta$ is the polar angle in the spherical coordinate system, and $\phi$ is the azimuthal angle in the same coordinate system. In this configuration, the photon momentum and polarization vectors can be expressed as
\begin{eqnarray}
	\hat{\mathbf{p}}&=&\sin(\theta_p)\cos(\phi_p)\,\hat{\mathbf{x}}+\sin(\theta_p)\sin(\phi_p)\,\hat{\mathbf{y}}+\cos(\theta_p)\,\hat{\mathbf{z}}~,\nonumber\\
	\hat{\bm{\varepsilon}}_1(\p)&=&\cos(\theta_p)\cos(\phi_p)\,\hat{\mathbf{x}}+\cos(\theta_p)\sin(\phi_p)\,\hat{\mathbf{y}}-\sin(\theta_p)\,\hat{\mathbf{z}}~,\nonumber\\
	\hat{\bm{\varepsilon}}_2(\p)&=&-\sin(\phi_p)\,\hat{\mathbf{x}}+\cos(\phi_p)\,\hat{\mathbf{y}}~.
\end{eqnarray}
Considering $\p$ in the direction of the $x$-axis, which means $\theta_p=\pi/2$, $\phi_p=0$, $\hat{\bm{\varepsilon}}_1(\k)=-\hat{\mathbf{z}}$, and $\hat{\bm{\varepsilon}}_2(\k)=\hat{\mathbf{y}}$ which results in
\begin{eqnarray} \label{eq:epsilon1dotB}
	\hat{\bm{\varepsilon}}_{1}(\k)\cdot\textbf{B}(\mathbf{r},t)&=&\frac{B_0}{2}\left(\frac{r_0}{r}\right)^3\hat{\bm{\varepsilon}}_{1}(\k)\cdot\big[3(\hat{\mathbf{m}}\cdot\hat{\mathbf{r}})\hat{\mathbf{r}}-\hat{\mathbf{m}}\big]\nonumber\\
	&=&\frac{B_0}{2}\left(\frac{r_0}{r}\right)^3\Big[-3\big(\cos(\theta)\cos(\theta_m)+\sin(\theta)\sin(\theta_m)\cos(\Omega t-\phi)\big)\cos(\theta)\nonumber\\
	&~&~~~~~~~~~~~~~~+\cos(\theta_m)\Big]~,
\end{eqnarray}
and
\begin{eqnarray}\label{eq:epsilon2dotB}
	\hat{\bm{\varepsilon}}_{2}(\k)\cdot\textbf{B}(\mathbf{r},t)&=&\frac{B_0}{2}\left(\frac{r_0}{r}\right)^3\hat{\bm{\varepsilon}}_{2}(\k)\cdot\big[3(\hat{\mathbf{m}}\cdot\hat{\mathbf{r}})\hat{\mathbf{r}}-\hat{\mathbf{m}}\big]\nonumber\\
	&=&\frac{B_0}{2}\left(\frac{r_0}{r}\right)^3\big[3\big(\cos(\theta)\cos(\theta_m)+\sin(\theta)\sin(\theta_m)\cos(\Omega t-\phi)\big)\sin(\theta)\sin(\phi)\nonumber\\
	&~&~~~~~~~~~~~~~~-\sin(\theta_m)\sin(\Omega t)\big]~.
\end{eqnarray}
Based on Eq.~\eqref{bg}, the magnetic field of an NS above the poles is strongest and directed towards $\hat{\mathbf{m}}$. On the equatorial plane it is weakest and directed towards $-\hat{\mathbf{m}}$. Therefore, we approximately consider $\mathbf{B}(\mathbf{r},t)$ within the interaction region as
\begin{eqnarray}\label{eq:magneticfield}
	\mathbf{B}(t)=B_\text{int}\hat{\mathbf{m}}~,
\end{eqnarray}
where $B_\text{int}=\frac{B_0r_0^3}{r_\text{int}^3}$ with $r_\text{int}$ being the radius at which the interaction takes place (see Fig.~\ref{fig:experiment}). Assuming an approximately constant interaction radius implies that the interaction time is smoothly limited.  Therefore, we use $\Pi(t/\tau)$ function denoting a smooth rectangular function that serves as
\begin{eqnarray}\label{eq:hevisidepi}
	\Pi\left(\frac{t}{\tau}\right)=
	\lim_{n \to \infty}\frac{1}{\left(\frac{2t}{\tau}\right)^n+1}~,
\end{eqnarray}
where $\tau$ is the time in which the interaction takes place. This time limit ensures that the photon experiences at least one magnetic field cycle while maintaining an approximately constant interaction radius. Under this assumption, the inner products of Eq.~\eqref{eq:epsilon1dotB} and Eq.~\eqref{eq:epsilon2dotB} reduce to
\begin{eqnarray} \label{eq:epsilondotB}
	\hat{\bm{\varepsilon}}_{1}(\k)\cdot\textbf{B}(\mathbf{r},t)&=&B_\text{int}\cos(\theta_m)\Pi(t/\tau)~,\nonumber\\
		\hat{\bm{\varepsilon}}_{2}(\k)\cdot\textbf{B}(\mathbf{r},t)&=&-B_\text{int}\sin(\theta_m)\sin(\Omega t)\Pi(t/\tau)~.
\end{eqnarray}
Upon substituting Eq.~\eqref{eq:epsilondotB} into the time evolution equations for Stokes parameters given in Eqs.~\eqref{eq:qdot}-\eqref{eq:vdot}, we obtain the following set of differential equations, which are derived in detail in Appendix~\ref{appendix:magnetic}.
\begin{eqnarray}\label{eq:stokedots}
	\dot{Q}(t)
	&\approx&-\frac{g^2_{a\gamma\gamma}B_{\text{int}}^2\omega_\k\sin(2\theta_m)}{4}
	\,\big[U(t)+iV(t)\big]D_F(\omega_{\k}+\Omega,\k)\sin(\Omega t)\Pi(t/\tau)
	~,\nonumber\\
	\dot{U}(t)&\approx&\frac{g^2_{a\gamma\gamma}B_{\text{int}}^2\omega_{\k}}{2 }\bigg[iV(t)\sin^2(\theta_m)\sin(\Omega t)+Q(t)\frac{\sin2\theta_m}{2}\bigg]D_F(\omega_{\k}+\Omega,\k)\sin(\Omega t)\Pi(t/\tau)~,\nonumber\\
	i  \dot{V}(t)	&\approx&\frac{g^2_{a\gamma\gamma}B_{\text{int}}^2\omega_{\k}}{2 }\bigg[U(t)\sin^2(\theta_m)\sin(\Omega t)-Q(t)\frac{\sin2\theta_m}{2}\bigg]D_F(\omega_{\k}+\Omega,\k)\sin(\Omega t)\Pi(t/\tau)~,\nonumber\\
\end{eqnarray}
where
\begin{eqnarray}\label{eq:propagator}
	D_F(\omega_{\k}+\Omega,\k)=\frac{1}{2\omega_{\k}\Omega+\Omega^2-m_a^2+i\gamma_B}~.
\end{eqnarray}
in which $\gamma_B=q^0\Gamma_B(q^0)$. 

As mentioned in Appendix~\ref{appendix:magnetic}, by setting the profile function poles (Eqs.~\eqref{P1}~and~\eqref{P2}) close to the propagator poles which means
\begin{eqnarray}\label{eq:NSresonance}
\Omega^2\pm2\Omega\omega_{\k}-m_a^2\approx0~,
\end{eqnarray}
an enhancement occurs. Therefore, for each ALP mass according to the rotation frequency of NS, the photon frequency at which it is most affected can be obtained by using Eq.~\eqref{eq:NSresonance}. The fastest known pulsar has a rotation frequency of $\Omega=716~\text{Hz}=4.7\times10^{-13}~\text{eV}$ \cite{hessels2006radio}. One condition is $\Omega^2+2\Omega\omega_{\k}-m_a^2\approx0$ and to have the interaction effect on positive values of $\omega_{\k}$ with the mentioned maximum rotation frequency,  ALP masses more than $4.7\times10^{-13}~\text{eV}$ should be studied. This lowest bound of $m_a$ decreases if the rotation frequency on NS is less than 716~Hz. On the other hand,  to have positive values of $\omega_{\k}$ based on the other condition, $\Omega^2-2\Omega\omega_{\k}-m_a^2\approx0$, ALP masses should be less than $4.7\times10^{-13}~\text{eV}$. Also, the photon frequency, $\omega_{\k}$, in this condition is less than radio frequencies and is hard to be detected. Therefore, the investigation with this approach covers ALP masses above $4.7\times10^{-13}~\text{eV}$ and use $\Omega^2+2\Omega\omega_{\k}-m^2\approx0$ as the main resonance condition.

There is an additional condition to be considered when approaching the resonance which involves examining the region of parameter space where the real values of the propagator in Eq.~\eqref{eq:propagator} are dominant that is $\Omega^2+2\omega_{\k}\Omega-m_a^2\gg\gamma_B$. Also, the enhancement effect occurs when the denominator of the propagator is much less than \sout{1} its numerator. Taking into account the above conditions means we should impose the following `Goldilocks' conditions on the parameter space
\begin{eqnarray}\label{eq:condition}
	\gamma_B\ll\Omega^2+2\omega_{\k}\Omega-m_a^2\ll\epsilon~.
\end{eqnarray}
where we consider $\epsilon$ to be about $1~\mathrm{eV}^2$. Therefore, in the following, we solve $\Omega^2+2\omega_{\k}\Omega-m_a^2=10\gamma_B$ for each value of $m_a$ and $g_{a\gamma\gamma}$ and examine it by Eq.~\eqref{eq:condition} to obtain the most affected photon frequency. Solving 
\begin{eqnarray}\label{eq:res}
	\Omega^2+2\omega_{\k}\Omega-m_a^2=10\gamma_B~,
\end{eqnarray}
based on Eq.~\eqref{barGmmm} to see the region of parameter space which gives positive values of $\omega_{\k}$ results the following theoretical bounds on the ALP mass 
\begin{eqnarray}\label{eq:mbound}
	m_a&\geq&\Omega~,
\end{eqnarray}
and the coupling constant
\begin{eqnarray}
	g_{a\gamma\gamma}\leq\frac{2\,\sqrt{2\Omega}}{B_{\text{int}}\sin\theta_m\sqrt{5\tau}}~.\label{eq:gbound}
\end{eqnarray}
Eq.~\eqref{eq:gbound} shows that there is an upper limit on the coupling constant in the exclusion areas. We see this bound in the results of the next section.


\section{Constraining ALPs Parameter Space: The Potential of the Scheme}\label{sec:scheme}

In this section, we utilize the insights obtained from the previously discussed results to determine the range of ALP mass and coupling constants that can be excluded. The observations indicate that neutron stars with higher frequencies have lower magnetic flux densities~\cite{Konar2010Magnetic}. However, the faster the NS spin, the greater the induced circular polarization, making millisecond pulsars more favorable options for observing this effect. To plot the sensitivity limit, we consider an NS with a magnetic field strength of $B_0=10^4~\text{T}$ and a rotation frequency of $\Omega=300~\text{Hz}$. We also assume that the photon experiences two cycles of the NS magnetic field. Therefore, the length at which the interaction takes place is $d=2c/\Omega\simeq2000~\text{km}$, where $c$ is the speed of light. The interaction occurs at a radius of $r_\text{int}=d=2000~\text{km}$, which approximately remains constant across $d$ as illustrated in  Fig.~\ref{fig:experiment} and we use $n=10$ in Eq.~\eqref{eq:hevisidepi} to ensure a smooth transition when applying the magnetic field. Additionally, as depicted in Fig.~\ref{fig:experiment} the misalignment angle of dipole should be almost $\pi/2$ therefore we consider $\theta_m=\pi/2$.

\begin{figure}[t]
	\includegraphics[width=\textwidth]{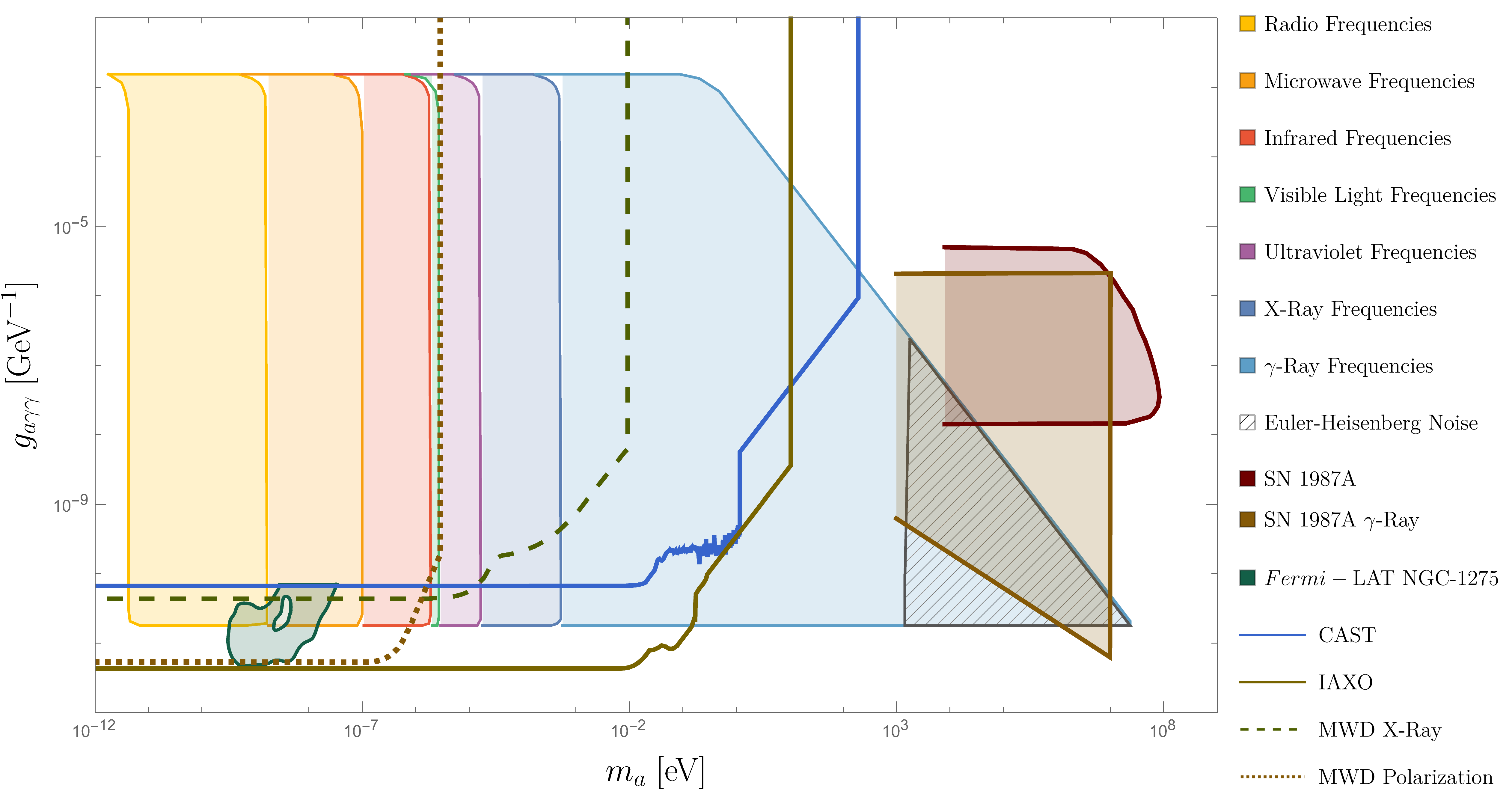}\centering
	\cprotect\caption{Expected limit on ALPs parameters for a NS with $B_0=10^4~\text{T}$, $\theta_m=\pi/2$, and $\Omega=300~\text{Hz}$. The plot illustrates the parameter space that can be excluded based on the observed circular polarization effect. Each color represents a distinct region of the electromagnetic spectrum that is affected, starting from the radio frequency range (yellow) on the left and progressing through microwave (orange), infrared (red), visible light (green), ultraviolet (purple), and X-ray (dark blue) before reaching the gamma-ray range (light blue) on the right. The hatched area at the end of the gamma-ray zone represents the region where the Euler-Heisenberg effect dominates the ALP effect in our study, as discussed in the Appendix~\ref{appendix:photonphoton}. Other experiments' data are provided from \verb|AxionLimits| repository \cite{AxionLimits}.}
	\label{fig:exc}
\end{figure}

To establish an exclusion region, we solve Eq.~\eqref{eq:res} to find the optimal value of $\omega_\mathbf{k}$ for each ALP mass and coupling constant. Next, we substitute the values of $m$, $g_{a\gamma\gamma}$, and the resulting $\omega_\mathbf{k}$ into Eq.~\eqref{eq:stokedots} and numerically solve it. The obtained value of $V(t)/I(t)$ at $t=\tau/2$ is then compared to the minimum detectable circular polarization $\left(V/I\right)_\text{min}=0.01$ \cite{perley2013integrated,bucciantini2018modeling}. If the value is discernible, the corresponding $m$ and $g_{a\gamma\gamma}$ are exported to construct the excluded region. The resulting excluded region is shown in Fig.~\ref{fig:exc}, in which we have categorized the obtained value of $\omega_\mathbf{k}$ for each $m$ and $g_{a\gamma\gamma}$ into seven different bands of the electromagnetic spectrum. In Fig.~\ref{fig:exc}, by moving from left to right the frequency of the photon that interacts the most, increases from radio frequency to gamma-ray. Regarding to the capability of detecting each frequency, a part of the parameter space could be explored. On the right side of the gamma-ray region, the hatched area represents the dominance of Euler-Heisenberg (EH) noise in our work. We look into the EH effect in Appendix~\ref{appendix:photonphoton} and show that, with the previously mentioned parameters, having $\omega_\k\leq2.1\times10^{32}~\text{Hz}$ guarantees a weaker EH induced circular polarizatioin compared to the second order ALP-photon scattering one. 

The resonance condition of Eq.~\eqref{eq:res} imposes an upper bound on the coupling constant based on Eq.~\eqref{eq:gbound} because large coupling constants results in negative values for $\omega_{\k}$ which are not acceptable. Additionally, we ensure that the value of $\gamma_B$ with the obtained $\omega_\mathbf{k}$ does not exceed $0.01\,\mathrm{eV}^2$ which creates a slope on the upper limit of coupling constant in large masses. Employing these procedures allows us to identify the parameter space satisfying the conditions of Eq.~\eqref{eq:condition}. Also, the minimum mass is theoretically limited to Eq.~\eqref{eq:mbound} and the maximum mass to the EH noise. Finally, as shown in Fig.~\ref{fig:exc} using numerical calculations this probe covers ALP masses in the region of $7\times10^{-12}~\text{eV}\leq m_a\leq1.5\times 10^{3}~\text{eV}$ with the coupling constants of $1.7\times10^{-11}~\text{GeV}^{-1}\leq g_{a\gamma\gamma}\leq1.6\times10^{-3}~\text{GeV}^{-1}$. This range of parameter space partially overlaps with other ALP experiments, especially the astronomical ones. We have chosen a modest value for the rotation frequency and the magnetic field strength so that includes a large number of millisecond pulsars. Also, we assumed the photon only senses two cycles of the magnetic field. Therefore, having a higher rotation frequency and magnetic field or assuming to sense more than two cycles improves the whole results to cover lower coupling constants. All of the numerical calculations have done utilizing Mathematica's \verb|NDSolve|. In Appendix~\ref{appendix:stokesanalytic} we analytically solved $\rho_{21}(t)$ and compared it to the numerical values of $U(t)+V(t)$ in order to validate the numerical calculation and it can be seen that numerical and analytical calculations are completely consistent. 

To observe the ALP-photon interaction effect, we propose studying a binary system comprising a NS and a companion star. The induced circular polarization from the interaction with ALPs can be observed when the companion star is positioned behind the NS relative to Earth. In this configuration, the effect becomes visible, but as the companion star moves in front of the NS, the effect diminishes, disappearing after half a period of the binary system. Among the hundreds of discovered millisecond NS systems, PSR J1748-2446ad stands out as a good candidate for observing the effect. This pulsar has a rotation frequency of $\Omega=716$~Hz which is 2.5 times higher than our presumed rotation frequency in plotting Fig~\ref{fig:exc}, a surface magnetic field of approximately $B_0\sim10^5~\text{T}$ which is ten times stronger than our assumption, and high eclipse fraction in its binary system (about 40\% of the orbit) \cite{hessels2006radio}. It forms a binary system with a companion star at a separation of $d_\text{b}=2.8\times10^6~\text{km}$, significantly longer than the expected interaction distance of $d=2000~\text{km}$. Additionally, the binary period is $26$~h, allowing ample time for investigating the circular polarization phenomenon. 


\section{Conclusion}

Our study delves into the phenomena arising from the interaction of photons with the rotating magnetic fields of NSs and their implications for constraining axion parameter space. Through our analysis, we demonstrate how the NS's magnetic field induces a transformation of previously linearly polarized photons into circularly polarized ones and the time periodicity of their magnetic field amplifies this interaction.

By proposing the utilization of binary systems composed of NSs and companion stars, we establish an effective means to observe the circular polarization phenomenon. Specifically, during partial eclipses of the companion star by the NS, the circular polarization becomes measurable due to the close proximity of the emitted photons to the NS's strong magnetic field. However, the effect diminishes as the companion star moves in front of the NS, which occurs after half a period of the binary system. Millisecond NSs are suitable candidates to probe the effect because of their high orbiting frequency and magnetic field. A millisecond NS with an orbital frequency of $\Omega=300~\text{Hz}$ and a magnetic field of $10^4$~T, excludes ALPs with couplings of $1.7\times10^{-11}~\text{GeV}^{-1}\leq g_{a\gamma\gamma}\leq1.6\times10^{-3}~\text{GeV}^{-1}$ and masses in the range of $7\times10^{-12}~\text{eV}\leq m_a\leq1.5\times 10^{3}~\text{eV}$.

Taking advantage of the unique characteristics of millisecond NSs, we identify PSR J1748-2446ad as an exemplary candidate for observing the circular polarization effect. Its favorable pulse period, surface magnetic field, and binary system parameters allow for extended observation periods, contributing to potential breakthroughs in studying ALP properties.

\section{Acknowledgment}

We express our sincere appreciation to  Nicholas L. Rodd for his valuable discussions and insightful comments, which greatly contributed to the improvement of this paper. Additionally, we would like to appreciate Marco Peloso for useful discussions and comments. M.~Z. would like to thank INFN and department of Physics and Astronomy “G. Galilei” at University of Padova and also the CERN Theory Division for warm hospitality while this work was done. 
\\
\appendix

\section{ALPs decay rate in the background of a time-dependent magnetic field}\label{appendix:decayrate}
In this appendix, we calculate the decay rate of axion in the presence of a time-dependent magnetic field. The amplitude for the axion-photon conversion process in the presence of the magnetic field is given by
\begin{eqnarray}
	\mathcal{T}_{fi}&=&\left<a,\mathbf{ q}\right|\frac{g_{ a\gamma\gamma}}{2}\int d^4x\, \epsilon^{\mu\nu\rho\sigma} \phi(x)\bar{F}_{\rho\sigma}(x)\partial_{\mu}A_{\nu}(x)\left|\gamma,\mathbf{k}\right>\\ \nonumber 
	&=&-\frac{ig_{\gamma a} \sqrt{\omega_{\mathbf{k}}}}{ V\sqrt{2q^0}}
	\int_{\tau/2}^{\tau/2} dt e^{-i(\omega_{\mathbf{k}}-q^0) t}
	\int_{V} d^3\x\,e^{-i(\q-\k)\cdot \x}\hat{\bm{\varepsilon}}(\mathbf{ k})\cdot\mathbf{B} (t)~.
\end{eqnarray}	
Using the assumptions of the main text ($\theta_p=\pi/2$, $\phi_p=0$, $\hat{\bm{\varepsilon}}_1(\k)=-\hat{\mathbf{z}}$, $\hat{\bm{\varepsilon}}_2(\k)=\hat{\mathbf{y}}$, and $\mathbf{B}(t)=-B_\text{int}\hat{\mathbf{m}}$) we have
\begin{eqnarray}
	\hat{\bm{\varepsilon}}_{1}(\k)\cdot\textbf{B}(\mathbf{r},t)=B_{\text{int}}\cos(\theta_m)~,
\end{eqnarray}
and
\begin{eqnarray}
	\hat{\bm{\varepsilon}}_{2}(\k)\cdot\textbf{B}(\mathbf{r},t)=-B_{\text{int}}\sin(\theta_m)\sin(\Omega t)\Pi(t/\tau)~.
\end{eqnarray}
Therefore, the conversion amplitude transforms to the form
\begin{eqnarray}
	\mathcal{T}_{fi}&=&-\frac{
		ig_{a\gamma\gamma} \sqrt{\omega_{\mathbf{k}}}B_{\text{int}}\sin\theta_m}{ V \sqrt{2q^0}}
	\int_{-\tau/2}^{\tau/2} dt\, e^{-i(\omega_{\mathbf{k}}-q^0) t}\sin\Omega t
	\int_{V} d^3\x e^{-i(\q-\k)\cdot \x}~.\nonumber\\
\end{eqnarray}
Here, we define the profile function
\begin{eqnarray} 
	P_1(q^0)=\int_{-\tau/2}^{\tau/2}  dt \sin(\Omega t)\,e^{-i(-q^0+\omega_{\mathbf{k}})t}=\frac{1}{i}\left(\frac{\sin{\left(\Delta_1\tau/2\right)}}{\Delta_1}-\frac{\sin{\left( \Delta_2 \tau/2\right)}}{\Delta_2}\right)~,
\end{eqnarray}
with $\Delta_1=q^0+\Omega-\omega_{\mathbf{k}}$, and $\Delta_2=q^0-\Omega-\omega_{\mathbf{k}}$.
The conversion probability per unit time is then given by
\begin{eqnarray}
	w=\frac{|\mathcal{T}_{fi}|^{2}}{\tau}=\frac{g^2_{ a\gamma\gamma}   \omega_{\mathbf{k}}B_{\text{int}}^2\sin^2\theta_m}{2V \tau q^0 } P_1^2(q^0)\,\,(2\pi)^3 \delta^3(\mathbf{q}-\mathbf{k})\, ~,
\end{eqnarray}
which leads to the following decay rate
\begin{eqnarray}\label{Gmmm}
	\Gamma_B(q^0)=\int w \frac{Vd^3\k}{(2\pi)^3}=\frac{g^2_{ a\gamma\gamma} \omega_\k B_{\text{int}}^2\sin^2\theta_m}{2q^0} \frac{P_1^2(q^0)}{\tau}~.
\end{eqnarray}
Approximating $P_1^2(q^0)/\tau\approx \tau/4$, we find a maximum value for the decay rate as
\begin{eqnarray} \label{barGmmm}
	\Gamma^{\textrm{max}}_B=\frac{g^2_{ a\gamma\gamma} \omega_\k B_{\text{int}}^2\sin^2\theta_m}{ 8\, q^0} \tau
	~,
\end{eqnarray}
which have been used in the main text.

\section{Time Evolution of the Stokes Parameters}\label{appendix:stoke}

Here we want to derive the time evolution of the Stokes parameters with the forward scattering term of QBE. In the QBE of Eq.~\eqref{boltzmann0}, the firs term on the right hand side is called the forward scattering term and the second term is called the collision term.  First, we insert the interaction Hamiltonian of Eq.~\eqref{H} into Eq.~\eqref{boltzmann0} and calculate the forward scattering term with the following expectation values as \cite{Zarei:2021dpb}
\begin{eqnarray} \label{EVP}
	\left<a_{s'}^{\dagger}(\p')a_{s}(\p)a^\dag_i(\mathbf{k}) a_j(\mathbf{k}) \right>_\textrm{c} &\simeq& 4p^0k^0(2\pi)^6\delta^3(\p'-\k)\delta^3(\p-\k)\,\delta^{si}\rho_{s'j}(\k)~, \\
	\left<a^\dag_i(\mathbf{k}) a_j(\mathbf{k})a_{s'}^{\dagger}(\p')a_{s}(\p) \right>_\textrm{c} &\simeq&4k^0p^0(2\pi)^6\delta^3(\k-\p)\delta^3(\k-\p')\, \delta^{js'}\rho_{is}(\p)~,
\end{eqnarray}
which leads to
\begin{eqnarray}
	i\l<\l[H_{\textrm{int}}(t),\hat{\mathcal{D}}_{ij}(\mathbf{k})\r]\r>&=&\frac{g^2_{a\gamma\gamma}}{2 }
	\,\sum_{s,s'}
	\int d^3rd^3r'dt' d\p\,d\p' \,\frac{d^4q}{(2\pi)^4}\omega_{\p}\omega_{\p'}D_F(q)\nonumber\\
	&\times&\left[\bm{\varepsilon}_{s}(\p)\cdot\textbf{B}(\mathbf{r},t)\right]\left[\bm{\varepsilon}_{s'}^*(\p')\cdot\textbf{B}(\mathbf{r}',t')\right]
	\nonumber \\ &\times & 
	\big[
	e^{-i(\q-\p)\cdot \mathbf{r}'}
	e^{-i (\p'-\q)\cdot   \mathbf{r}}
	e^{-i(-q^0+\omega_{\mathbf{p}})t'}
	e^{-i(q^0-\omega_{\mathbf{p}'})t}\nonumber\\
	&+&e^{-i(\q+\p')\cdot \mathbf{r}'}
	e^{-i(-\q-\p)\cdot \mathbf{r} }
	e^{-i(-q^0-\omega_{\p'})t'}
	e^{-i(q^0+\omega_{\p})t}
	\big]\nonumber\\
	&\times&4\,p^0k^0(2\pi)^6\delta^3(\k-\p)\delta^3(\k-\p')\Big[\delta^{si}\rho_{s'j}(\k)-\, \delta^{js'}\rho_{is}(\p)\Big]~.\nonumber\\
\end{eqnarray}
Now taking the integration over $\p$ and $\p'$ we obtain
\begin{eqnarray}
	i\l<\l[H_{\textrm{int}}(t),\hat{\mathcal{D}}_{ij}(\mathbf{k})\r]\r>&=&\frac{g^2_{a\gamma\gamma}}{2 }
	\,\sum_{s,s'}
	\int d^3rd^3r'dt'  \,\frac{d^4q}{(2\pi)^4}\omega_{\k}^2D_F(q)
	\left[\bm{\varepsilon}_{s}(\k)\cdot\textbf{B}(\mathbf{r},t)\right]\left[\bm{\varepsilon}_{s'}^*(\k)\cdot\textbf{B}(\mathbf{r}',t')\right]
	\nonumber \\ &\times & 
	\Big[
	e^{-i(\q-\k)\cdot \mathbf{r}'}
	e^{-i (\k-\q)\cdot   \mathbf{r}}
	e^{-i(-q^0+\omega_{\mathbf{k}})t'}
	e^{-i(q^0-\omega_{\mathbf{k}'})t}\nonumber\\
	&~&+e^{-i(\q+\k)\cdot \mathbf{r}'}
	e^{-i(-\q-\k)\cdot \mathbf{r} }
	e^{-i(-q^0-\omega_{\k})t'}
	e^{-i(q^0+\omega_{\k})t}
	\Big]\Big[\delta^{si}\rho_{s'j}(\k)-\, \delta^{js'}\rho_{is}(\k)\Big]~.\nonumber\\
\end{eqnarray}
Summing over the polarizations leads to constructing the following time evolution of Stokes parameters
\begin{eqnarray}
	&&(2\pi)^3 \delta^3(0)\dot{Q}(t)=(2\pi)^3 \delta^3(0)(\dot{\rho}_{11}-\dot{\rho}_{22})\nonumber\\
	&=&\frac{g^2_{a\gamma\gamma}}{2 }
	\,
	\int d^3rd^3r'dt'  \,\frac{d^4q}{(2\pi)^4}\,\omega_{\k}\,D_F(q)
	\nonumber \\ &\times & 
	\left[
	e^{-i(\q-\k)\cdot \mathbf{r}'}
	e^{-i (\k-\q)\cdot   \mathbf{r}}
	e^{-i(-q^0+\omega_{\mathbf{k}})t'}
	e^{-i(q^0-\omega_{\mathbf{k}'})t}+e^{-i(\q+\k)\cdot \mathbf{r}'}
	e^{-i(-\q-\k)\cdot \mathbf{r} }
	e^{-i(-q^0-\omega_{\k})t'}
	e^{-i(q^0+\omega_{\k})t}
	\right]\nonumber\\
	&\times&\Bigg[U(t)\Big(\left[\bm{\varepsilon}_{1}(\k)\cdot\textbf{B}(\mathbf{r},t)\right]
	\left[\bm{\varepsilon}_2^*(\k)\cdot\textbf{B}(\mathbf{r}',t')\right]-\left[\bm{\varepsilon}_2(\k)\cdot\textbf{B}(\mathbf{r},t)\right]\left[\bm{\varepsilon}^*_{1}(\k)\cdot\textbf{B}(\mathbf{r}',t')\right]
	\Big)\nonumber\\
	&+&iV(t)\Big(\left[\bm{\varepsilon}_{1}(\k)\cdot\textbf{B}(\mathbf{r},t)\right]
	\left[\bm{\varepsilon}_2^*(\k)\cdot\textbf{B}(\mathbf{r}',t')\right]+
	\left[\bm{\varepsilon}_2(\k)\cdot\textbf{B}(\mathbf{r},t)\right]\left[\bm{\varepsilon}^*_{1}(\k)\cdot\textbf{B}(\mathbf{r}',t')\right]\Big)\Bigg]~,
\end{eqnarray}
\begin{eqnarray}
	&&(2\pi)^3 \delta^3(0)\dot{U}(t)=(2\pi)^3 \delta^3(0)(\dot{\rho}_{12}+\dot{\rho}_{21})\nonumber\\
	&=&\frac{g^2_{a\gamma\gamma}}{2 }
	\,
	\int d^3rd^3r'dt'  \,\frac{d^4q}{(2\pi)^4}\,\omega_{\k}D_F(q)
	\nonumber \\ &\times & 
	\left[
	e^{-i(\q-\k)\cdot \mathbf{r}'}
	e^{-i (\k-\q)\cdot   \mathbf{r}}
	e^{-i(-q^0+\omega_{\mathbf{k}})t'}
	e^{-i(q^0-\omega_{\mathbf{k}'})t}+e^{-i(\q+\k)\cdot \mathbf{r}'}
	e^{-i(-\q-\k)\cdot \mathbf{r} }
	e^{-i(-q^0-\omega_{\k})t'}
	e^{-i(q^0+\omega_{\k})t}
	\right]\nonumber\\
	&\times&\Bigg[iV(t)\Big(-\left[\bm{\varepsilon}_{1}(\k)\cdot\textbf{B}(\mathbf{r},t)\right]
	\left[\bm{\varepsilon}_1^*(\k)\cdot\textbf{B}(\mathbf{r}',t')\right]+\left[\bm{\varepsilon}_{2}(\k)\cdot\textbf{B}(\mathbf{r},t)\right]
	\left[\bm{\varepsilon}_2^*(\k)\cdot\textbf{B}(\mathbf{r}',t')\right]\Big)\nonumber\\
	&+&Q(t)\Big(-\left[\bm{\varepsilon}_{1}(\k)\cdot\textbf{B}(\mathbf{r},t)\right]
	\left[\bm{\varepsilon}_2^*(\k)\cdot\textbf{B}(\mathbf{r}',t')\right]+\left[\bm{\varepsilon}_{2}(\k)\cdot\textbf{B}(\mathbf{r},t)\right]
	\left[\bm{\varepsilon}_1^*(\k)\cdot\textbf{B}(\mathbf{r}',t')\right]\Big)\Bigg]~,\nonumber\\
\end{eqnarray}
and
\begin{eqnarray}
	&&i(2\pi)^3 \delta^3(0)\dot{V}(t)=(2\pi)^3 \delta^3(0)(\dot{\rho}_{21}-\dot{\rho}_{12})\nonumber\\
	&=&\frac{g^2_{a\gamma\gamma}}{2 }
	\,
	\int d^3rd^3r'dt'  \,\frac{d^4q}{(2\pi)^4}\,\omega_{\k}\,D_F(q)
	\nonumber \\ &\times & 
	\left[
	e^{-i(\q-\k)\cdot \mathbf{r}'}
	e^{-i (\k-\q)\cdot   \mathbf{r}}
	e^{-i(-q^0+\omega_{\mathbf{k}})t'}
	e^{-i(q^0-\omega_{\mathbf{k}'})t}+e^{-i(\q+\k)\cdot \mathbf{r}'}
	e^{-i(-\q-\k)\cdot \mathbf{r} }
	e^{-i(-q^0-\omega_{\k})t'}
	e^{-i(q^0+\omega_{\k})t}
	\right]\nonumber\\
	&\times&\Bigg[U(t)\Big(-\left[\bm{\varepsilon}_{1}(\k)\cdot\textbf{B}(\mathbf{r},t)\right]
	\left[\bm{\varepsilon}_1^*(\k)\cdot\textbf{B}(\mathbf{r}',t')\right]+\left[\bm{\varepsilon}_{2}(\k)\cdot\textbf{B}(\mathbf{r},t)\right]
	\left[\bm{\varepsilon}_2^*(\k)\cdot\textbf{B}(\mathbf{r}',t')\right]\Big)\nonumber\\
	&+&Q(t)\Big(\left[\bm{\varepsilon}_{1}(\k)\cdot\textbf{B}(\mathbf{r},t)\right]
	\left[\bm{\varepsilon}_2^*(\k)\cdot\textbf{B}(\mathbf{r}',t')\right]+\left[\bm{\varepsilon}_{2}(\k)\cdot\textbf{B}(\mathbf{r},t)\right]
	\left[\bm{\varepsilon}_1^*(\k)\cdot\textbf{B}(\mathbf{r}',t')\right]\Big)\Bigg]~.\nonumber\\
\end{eqnarray}
Having the magnetic field and the photon polarizations and using them inside the above differential equations gives the time evolustion of Stokes parameters.

\section{Stokes Parameters with the NS Magnetic Field}\label{appendix:magnetic}
This appendix is devoted to obtain the time evolution of Stokes parameters using Eq.~\eqref{eq:magneticfield} as the magnetic field of the NS. We use the magnetic field of Eq.~\eqref{eq:epsilondotB} in Eqs.~\eqref{eq:qdot}-\eqref{eq:vdot} and obtain
\begin{eqnarray}
	\dot{Q}(t)&=&\frac{g^2_{a\gamma\gamma}B_{\text{int}}^2\omega_{\k}}{2}
	\int dt'  \,\frac{dq^0}{(2\pi)}D_F(q^0,\k)\left[
	e^{-i(-q^0+\omega_{\mathbf{k}})t'}
	e^{-i(q^0-\omega_{\mathbf{k}})t}+
	e^{-i(-q^0-\omega_{\k})t'}
	e^{-i(q^0+\omega_{\k})t}
	\right]\nonumber\\
	&\times&\Bigg[U(t)\frac{\sin(2\theta_m)}{2}\Big(-\sin(\Omega t')+\sin(\Omega t)
	\Big)\nonumber\\
	&-&iV(t)\frac{\sin(2\theta_m)}{2}\Big(\sin(\Omega t')+
	\sin(\Omega t)\Big)\Bigg]\Pi(t/\tau)\Pi(t'/\tau)~,
\end{eqnarray}
\begin{eqnarray}
	\dot{U}(t)&=&\frac{g^2_{a\gamma\gamma}B_{\text{int}}^2\omega_{\k}}{2 }
	\,
	\int dt'  \,\frac{dq^0}{(2\pi)}\,D_F(q^0,\k)
	\left[
	e^{-i(-q^0+\omega_{\mathbf{k}})t'}
	e^{-i(q^0-\omega_{\mathbf{k}})t}+
	e^{-i(-q^0-\omega_{\k})t'}
	e^{-i(q^0+\omega_{\k})t}
	\right]\nonumber\\
	&\times&\Bigg[iV(t)\Big(-\cos^2(\theta_m)+\sin^2(\theta_m)\sin(\Omega t)\sin(\Omega t')\Big)\nonumber\\
	&+&Q(t)\frac{\sin(2\theta_m)}{2}\Big(
	\sin(\Omega t')-\sin(\Omega t)\Big)\Bigg]\Pi(t/\tau)\Pi(t'/\tau)~,\nonumber\\
\end{eqnarray}
and
\begin{eqnarray}
	i \dot{V}(t)&=&\frac{g^2_{a\gamma\gamma}B_{\text{int}}^2\omega_{\k}}{2 }
	\,
	\int dt'  \,\frac{dq^0}{(2\pi)}\,D_F(q^0,\k)
	\left[
	e^{-i(-q^0+\omega_{\mathbf{k}})t'}
	e^{-i(q^0-\omega_{\mathbf{k}'})t}+
	e^{-i(-q^0-\omega_{\k})t'}
	e^{-i(q^0+\omega_{\k})t}
	\right]\nonumber\\
	&\times&\Bigg[U(t)\Big(-\cos^2(\theta_m)+\sin^2(\theta_m)\sin(\Omega t)	\sin(\Omega t')\Big)\nonumber\\
	&-&Q(t)\frac{\sin(2\theta_m)}{2}\Big(
	\sin(\Omega t')+\sin(\Omega t)\Big)\Bigg]\Pi(t/\tau)\Pi(t'/\tau)~.\nonumber\\
\end{eqnarray}
We start from $Q$ parameter and integrate over $t'$ which leads to
\begin{eqnarray} \label{eq:qdotcalc0}
	\dot{Q}(t)&=&\frac{g^2_{a\gamma\gamma}B_{\text{int}}^2\omega_\k\sin(2\theta_m)\Pi(t/\tau)}{4}
	\,\int\frac{dq^0}{(2\pi)}\Bigg\{\big[U(t)-iV(t)\big]D_F(q^0,\k)\sin(\Omega t) \nonumber\\
	&\times&\bigg[2\pi \delta(q^0-\omega_{\k})
	e^{-i(q^0-\omega_{\mathbf{k}})t}+2\pi \delta(q^0+\omega_{\k})e^{-i(q^0+\omega_{\k})t}\bigg]\nonumber\\
	&-& \,D_F(q^0,\k)\big[U(t)+iV(t)\big]\left[P_1(q^0)
	e^{-i(q^0-\omega_{\mathbf{k}})t}+P_2(q^0)e^{-i(q^0+\omega_{\k})t}\right]\Bigg\}~,\nonumber\\
\end{eqnarray}
where
\begin{eqnarray} \label{P1}
	P_1(q^0)=\int_{-\infty}^{\infty}  dt' \sin(\Omega t')\Pi(t'/\tau)\,e^{-i(-q^0+\omega_{\mathbf{k}})t'}=\frac{1}{i}\left(\frac{\sin{\left(\Delta_1\tau/2\right)}}{\Delta_1}-\frac{\sin{\left( \Delta_2 \tau/2\right)}}{\Delta_2}\right)~,\nonumber\\
\end{eqnarray}
and
\begin{eqnarray}  \label{P2}
	P_2(q^0)=\int_{-\infty}^{\infty}  dt' \sin(\Omega t')\Pi(t'/\tau)\,e^{-i(-q^0-\omega_{\mathbf{k}})t'}=\frac{1}{i}\left(\frac{\sin{\left(\Delta_3\tau/2\right)}}{\Delta_3}-\frac{\sin{\left( \Delta_4 \tau/2\right)}}{\Delta_4}\right)~,\nonumber\\
\end{eqnarray}
with $\Delta_1=q^0+\Omega-\omega_{\mathbf{k}}$, $\Delta_2=q^0-\Omega-\omega_{\mathbf{k}}$
, $\Delta_3=q^0+\Omega+\omega_{\mathbf{k}}$, and $\Delta_4=q^0-\Omega+\omega_{\mathbf{k}}$. Under the condition of $\omega_{\k}\tau\gg1$ which is true in the scenarios of this manuscript, one can replace the sinc functions of the above profiles with Dirac delta functions as  $\frac{\sin{\left( \Delta_i \tau/2\right)}}{\Delta_i}=\pi\delta(\Delta_i)$. The propagator in Eq.~\eqref{eq:qdotcalc0} exhibits four poles at $q^0= \pm\sqrt{\omega_\mathbf{k}^2-m_a^2 \pm \gamma^2_B }$, while the profile functions, based on Eqs.~\eqref{P1} and \eqref{P2}, have four poles at $q^0=\pm\omega_{\k}\pm\Omega$. The maximum enhancement occures when the propagator poles and profile poles get close te eachother. In other words, the real part of the propagator's denominator approaches to zero that results the resonance conditions as
\begin{eqnarray}
\Omega^2\pm2\Omega\omega_{\k}-m_a^2\approx0,
\end{eqnarray}
which makes and enhancement effect in the third line of Eq.~\eqref{eq:qdotcalc0}. Replacing the sinc functions with Dirac delta functions in Eq.~\eqref{eq:qdotcalc0} we obtain
\begin{eqnarray}\label{eq:qdotcalc1}
	\dot{Q}(t)&=&\frac{g^2_{a\gamma\gamma}B_{\text{int}}^2\omega_\k\sin(2\theta_m)\Pi(t/\tau)}{4}
	\,\int\frac{dq^0}{(2\pi)}\Bigg\{\big[U(t)-iV(t)\big]D_F(q^0,\k)\sin(\Omega t) \nonumber\\
	&\times&\bigg[2\pi \delta(q^0-\omega_{\k})
	e^{-i(q^0-\omega_{\mathbf{k}})t}+2\pi \delta(q^0+\omega_{\k})e^{-i(q^0+\omega_{\k})t}\bigg]\nonumber\\
	&-& \,D_F(q^0,\k)\big[U(t)+iV(t)\big]\nonumber\\
	&\times&\frac{1}{i}\left[(\pi \delta(\Delta_1)-\pi \delta(\Delta_2))
	e^{-i(q^0-\omega_{\mathbf{k}})t}+(\pi \delta(\Delta_3)-\pi \delta(\Delta_4))e^{-i(q^0+\omega_{\k})t}\right]\Bigg\}~.\nonumber\\
\end{eqnarray}
Taking the integration over $q^0$ results in
\begin{eqnarray}\label{eq:qdotcalc2}
	\dot{Q}(t)&=&\frac{g^2_{a\gamma\gamma}B_{\text{int}}^2\omega_\k\sin(2\theta_m)}{4}
	\,\Bigg\{2\big[U(t)-iV(t)\big]D_F(\omega_{\k},\k)\sin(\Omega t)\nonumber\\
	&-& \big[U(t)+iV(t)\big]\Big[D_F(\omega_{\k}-\Omega,\k)+D_F(\omega_{\k}+\Omega,\k)\Big]\sin(\Omega t)\Bigg\}\Pi(t/\tau)\nonumber\\
	&\approx&-\frac{g^2_{a\gamma\gamma}B_{\text{int}}^2\omega_\k\sin(2\theta_m)}{4}\sin2\theta_m
	\,\big[U(t)+iV(t)\big]D_F(\omega_{\k}+\Omega,\k)\sin(\Omega t)\Pi(t/\tau)~,\nonumber\\
\end{eqnarray}
where
\begin{eqnarray}
	D_F(\omega_{\k}+\Omega,\k)=\frac{1}{2\omega_{\k}\Omega+\Omega^2-m_a^2+i\gamma_B},
\end{eqnarray}
with $\gamma_B=q^0\Gamma_B(q^0)$. In the last line approximation of Eq.~\eqref{eq:qdotcalc2} we chose only those profile poles which result a positive value of $\omega_{\k}$ when overlap with the propagator poles. Using the same procedure for $U$ and $V$ parameters we have
\begin{eqnarray}
	 \dot{U}(t)&=&\frac{g^2_{a\gamma\gamma}B_{\text{int}}^2\omega_{\k}}{2 }
	\,\Bigg\{2D_F(\omega_{\mathbf{k}},\k)\left(-iV(t)\cos^2(\theta_m)-Q(t)\frac{\sin(2\theta_m)}{2}\sin(\Omega t)\right)\nonumber\\
	&+& \,\int\frac{dq^0}{(2\pi)}\,D_F(q^0,\k)\Bigg[iV(t)\sin^2(\theta_m)\sin(\Omega t)+Q(t)\frac{\sin(2\theta_m)}{2}\Bigg]\nonumber\\
	&\times&\left[P_1(q^0)
	e^{-i(q^0-\omega_{\mathbf{k}})t}+P_2(q^0)e^{-i(q^0+\omega_{\k})t}\right]\Bigg\}\Pi(t/\tau)\nonumber\\
	\nonumber\\
	&\approx&\frac{g^2_{a\gamma\gamma}B_{\text{int}}^2\omega_{\k}}{2 }\bigg[iV(t)\sin^2(\theta_m)\sin(\Omega t)+Q(t)\frac{\sin2\theta_m}{2}\bigg]D_F(\omega_{\k}+\Omega,\k)\sin(\Omega t)\Pi(t/\tau)~,\nonumber\\
\end{eqnarray}
and
\begin{eqnarray}
	i \dot{V}(t)&=&\frac{g^2_{a\gamma\gamma}B_{\text{int}}^2\omega_{\k}}{2 }
	\,\Bigg\{2D_F(\omega_{\mathbf{k}},\k)\left(-U(t)\cos^2(\theta_m)-Q(t)\frac{\sin(2\theta_m)}{2}\sin(\Omega t)\right)\nonumber\\
	&+& \,\int\frac{dq^0}{(2\pi)}\,D_F(q^0,\k)\Bigg[U(t)\sin^2(\theta_m)\sin(\Omega t)-Q(t)\frac{\sin(2\theta_m)}{2}\Bigg]\nonumber\\
	&\times&\left[P_1(q^0)
	e^{-i(q^0-\omega_{\mathbf{k}})t}+P_2(q^0)e^{-i(q^0+\omega_{\k})t}\right]\Bigg\}\Pi(t/\tau)	\nonumber\\
	&\approx&\frac{g^2_{a\gamma\gamma}B_{\text{int}}^2\omega_{\k}}{2 }\bigg[U(t)\sin^2(\theta_m)\sin(\Omega t)-Q(t)\frac{\sin2\theta_m}{2}\bigg]D_F(\omega_{\k}+\Omega,\k)\sin(\Omega t)\Pi(t/\tau)~.\nonumber\\
\end{eqnarray}
We use the above differential equations in the main text to probe the effect of the interaction on the Stokes parameters of the photon.

\section{Photon-Photon Scattering Effect}\label{appendix:photonphoton}
In the strong magnetic fields the Euler-Heisenberg (EH) effect~\cite{euler1935scattering} which is the photon-photon scattering in the context of quantum electrodynamics (QED) should be took into account. We start from EH Lagrangian density \cite{raffelt1988mixing}
\begin{eqnarray}
	\mathcal{L}_{QED}=\frac{\alpha^2}{90m_e^4}\left[(F_{\mu\nu}F^{\mu\nu})^2+\frac{7}{4}(F_{\mu\nu}\tilde{F}^{\mu\nu})^2\right]~,
\end{eqnarray}
in which $\alpha=\frac{1}{137}$ is the fine structure constant and $m_e$ is the electron mass. By splitting the electromagnetic tensor into a homogeneous part and a fluctuating one the Lagrangian density takes the form of
\begin{eqnarray}
	\mathcal{L}_{QED}=\frac{\alpha^2}{90m_e^4}\left[16\left(\bar{F}_{\mu\nu}\,\partial^\mu A^\nu\right)^2+7\left(\epsilon^{\mu\nu\rho\sigma}\bar{F}_{\mu\nu}\,\partial_\rho A_\sigma\right)^2\right]~,
\end{eqnarray}
where $\epsilon^{\mu\nu\rho\sigma}$ is the Levi-Civita tensor. The resulting interaction Hamiltonian density for the first order contribution is
\begin{eqnarray}
	\mathcal{H}_{\text{int}}=\frac{\alpha^2}{90m_e^4}\left[16\bar{F}^{\mu'\nu'}\,\partial_{\mu'} A^-_{\nu'}(x)\bar{F}^{\mu\nu}\,\partial_\mu A^+_\nu(x)+7\epsilon^{\mu'\nu'\rho'\sigma'}\bar{F}_{\mu'\nu'}\,\partial_{\rho'} A^-_{\sigma'}(x)\epsilon^{\mu\nu\rho\sigma}\tilde{F}_{\mu\nu}\,\partial_\rho A^+_\sigma(x)\right]~.\nonumber\\
\end{eqnarray}
Using Eq.~\eqref{Amu} as the photon field we have
\begin{eqnarray}
	&&\bar{F}^{\mu\nu}\,\partial_\mu A^+_\nu(x)\simeq\bar{F}^{0j}\,\partial_0 A^+_j(x)=\sum_s\int d\p\frac{i}{2p^0}\left[\textbf{B}(\textbf{r},t).(\p\times\bm{\varepsilon}_{s}(\p))\right]a_s(\p)e^{ip\cdot x}~,\nonumber\\
	&&\epsilon^{\mu\nu\rho\sigma}\tilde{F}_{\mu\nu}\,\partial_\rho A^+_\sigma(x)\simeq B_j\,\partial_0 A^+_j(x)=\sum_s\int d\p\frac{i}{2}\left[\textbf{B}(\textbf{r},t).\bm{\varepsilon}_{s}(\p)\right]a_s(\p)e^{ip\cdot x}~.
\end{eqnarray}
The interaction Hamiltonian then reads
\begin{eqnarray}
	H_{\text{int}}&=&\frac{\alpha^2}{360m_e^4}\sum_{s,s'}\int d^3r d\p d\p'  \Big\{16\,\, \big[\textbf{B}'(\textbf{r},t).(\hat{\p}'\times\bm{\varepsilon}_{s'}(\p'))\big]\,\,\big[\textbf{B}(\textbf{r},t).(\hat{\p}\times\bm{\varepsilon}_{s}(\p))\big]\nonumber\\
	&+&7[\textbf{B}'(\textbf{r},t).\bm{\varepsilon}_{s'}(\p')]\,\,[\textbf{B}(\textbf{r},t).\bm{\varepsilon}_{s}(\p)]\Big\}a_{s'}^\dagger(\p')a_s(\p)e^{-i(p'-p)\cdot x}\nonumber\\
	&=&\frac{\alpha^2}{360m_e^4}\sum_{s,s'}\int  d\p  \Big\{16 \big[\textbf{B}'(t).(\hat{\p}\times\bm{\varepsilon}_{s'}(\p))\big]\big[\textbf{B}(t).(\hat{\p}\times\bm{\varepsilon}_{s}(\p))\big]\nonumber\\
	&+&7[\textbf{B}'(t).\bm{\varepsilon}_{s'}(\p)][\textbf{B}(t).\bm{\varepsilon}_{s}(\p)]\Big\}a_{s'}^\dagger(\p)a_s(\p)~,
\end{eqnarray}
in which at the last line it is assumed to have a constant radius for the magnetic field in the distance of interaction as done in Appendix \ref{appendix:stoke}. Using the forward scattering term of QBE and the expectation values of Eqs.~\eqref{EVP}
\begin{eqnarray}
	\l<\l[H_{\textrm{int}}(t),\hat{\mathcal{D}}_{ij}(\mathbf{k})\r]\r>&=&\frac{\alpha^2}{90m_e^4}\sum_{s,s'}\int  d\p\, \mathcal{B}(t)\,p^0\,k^0\left(\delta^{si}\rho_{s'j}(\k)-\delta^{js'}\rho_{is}(\p)\right)\nonumber\\
	&\times&(2\pi)^6\delta^3(\p-\k)\delta^3(\p-\k)\nonumber\\
	&=&\frac{\alpha^2\omega_{\k}^2}{90m_e^4}\sum_{s,s'} \mathcal{B}(t)\left(\delta^{si}\rho_{s'j}(\k)-\delta^{js'}\rho_{is}(\k)\right)(2\pi)^3\delta^3(0)~,
\end{eqnarray}
where
\begin{eqnarray}
	\mathcal{B}(t)=16 \big[\textbf{B}'(t).(\hat{\p}\times\bm{\varepsilon}_{s'}(\p))\big]\big[\textbf{B}(t).(\hat{\p}\times\bm{\varepsilon}_{s}(\p))\big]+7[\textbf{B}'(t).\bm{\varepsilon}_{s'}(\p)][\textbf{B}(t).\bm{\varepsilon}_{s}(\p)]~,
\end{eqnarray}
is in the order of $B_{\text{int}}^2$. 
Now we can compare the effect of ALP-photon interaction near the NS with its EH counterpart. The ALP-photon interaction is dominant if
\begin{eqnarray}\label{eq:dom}
	\frac{g_{a\gamma\gamma}^2\,B_{\text{int}}^2\,\omega_{\k}}{2\omega_{\k}\Omega+\Omega^2-m^2}\geq\frac{\alpha^2\omega_{\k}B_{\text{int}}^2}{90\,m_e^4}~.
\end{eqnarray}
As discussed in Section~\ref{sec:scheme}, the photons used for the exclusion area are the solution of $\Omega^2+2\omega_{\k}\Omega-m_a^2=10\gamma_B$. Therefore, the ALP dominance condition of Eq.~\eqref{eq:dom} leads to
\begin{eqnarray}
	\omega_{\k}\leq\frac{72\,m_e^4}{B_{\text{int}}^2\tau\alpha^2\sin^2\theta_m}~.
\end{eqnarray}
Using parameters of Section~\ref{sec:scheme} which are $\theta_m=\pi/2$, $\tau=6.7$ ms, and $B_0=10^4$ T means an upper limit of $\omega_{\k}\leq2.1\times10^{32}~\text{Hz}$ for the photons. Thus, for less frequencies the EH induced circular polarization is subdominant and the ALP induced one is stronger. The region of dominant EH noise in our work is shown by hatched area in Fig.~\ref{fig:exc}. Notably, the highest frequency associated with the rightest edge of our exclusion region is $\omega_{\k}=8.4\times10^{41}~\text{Hz}$. This is an order of magnitude lower than the Planck frequency, $\omega_{\mathbf{P}}=1.87\times10^{43}~\text{Hz}$.\\
\\

\section{Validation of Stokes Parameters' Numerical Calculation}\label{appendix:stokesanalytic}

Here, we want to validate the numerical calculations with an analytical expression. Starting from Eq.~\eqref{eq:stokedots} and combining the second and the third relations results in
\begin{eqnarray}\label{eq:rhodot}
	\dot{\rho}_{21}(t)\approx\frac{1}{2} g^2_{a\gamma\gamma}B_{\text{int}}^2\omega_\k\,\rho_{21}(t)\sin^2(\theta_m)\sin^2(\Omega t)\Pi(t/\tau)D_F(\omega_{\k}+\Omega,\k)~.
\end{eqnarray}

This differential equation could be solved analytically, which yields the following piecewise function
\begin{eqnarray}\label{eq:rho}
	\rho_{21}(t)=
	\begin{cases} 
		0, & t\leq -\frac{\tau}{2} \\
		\exp\left[\frac{1}{2}g^2_{a\gamma\gamma}B_{\text{int}}^2\omega_\k\,\sin^2(\theta_m)\left(\frac{\Omega(2t+\tau)-\sin(2t\Omega)-\sin(\tau\Omega)}{4\Omega}\right)D_F(\omega_{\k}+\Omega,\k)\right], & -\frac{\tau}{2}\leq t\leq \frac{\tau}{2} \\
		\exp\left[\frac{1}{2}g^2_{a\gamma\gamma}B_{\text{int}}^2\omega_\k\,\sin^2(\theta_m)\left(\frac{\Omega\tau-\sin(\tau\Omega)}{\Omega}\right)D_F(\omega_{\k}+\Omega,\k)\right]. & \frac{\tau}{2}\leq t
	\end{cases}	\nonumber\\
\end{eqnarray}

\begin{figure}[t]
	\includegraphics[width=.98\textwidth]{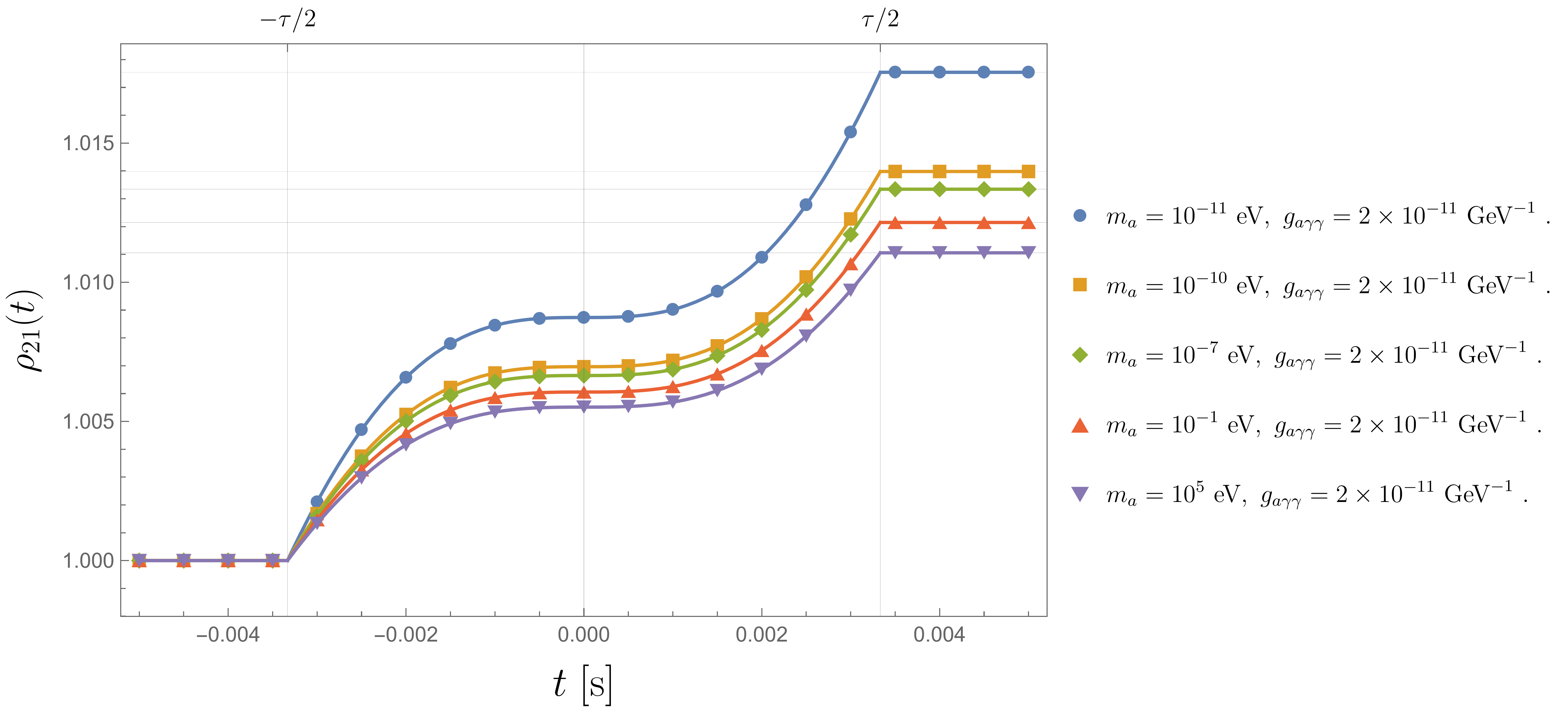}\centering
	\caption{A comparison between the analytical $\rho_{21}(t)$ in Eq.~\eqref{eq:rho} (solid lines) and the numerical one from Eq.~\eqref{eq:stokedots} (markers) for different masses. The mass and the coupling constants are mentioned next to the figure, the frequency is obtained by the resonance condition, and other parameters are the same as those for  Fig.~\ref{fig:exc}. Stokes parameters start out with initial values of  $U(-\tau/2)=1$, and $Q(-\tau/2)=V(-\tau/2)=0$. The numerical calculation is in a good agreement with the analytic one.}
	\label{fig:nuan}
\end{figure}

Since the Eqs.~\eqref{eq:stokedots} cannot be solved analytically for each Stokes parameter, we employ numerical methods for their solution. In order to validate the results, we utilize the analytical expression of $\rho_{21}(t)$ in Eq.~\eqref{eq:rho} and compare it to numerical solving of Eqs.~\eqref{eq:stokedots}. The outcome is displayed in Fig.~\ref{fig:nuan}, demonstrating the great accuracy of the numerical calculation. The parameters used for producing Fig.~\ref{fig:nuan} are the same as those used in Fig.~\ref{fig:exc} for some masses mentioned in the caption and the photon frequency is on the resonance condition of $\Omega^2+2\omega_{\k}\Omega-m_a^2=10\gamma_B$ like Fig.~\ref{fig:nuan}.
\bibliography{references}

\end{document}